\documentclass[aps,prd,groupedaddress,showpacs,twocolumn]{revtex4}
\usepackage{epsf,times}
\def\np#1#2#3   {{ Nucl. Phys.} {\bf#1}, #2 (#3). }
\def\pcps#1#2#3 {{ Proc. Cam. Phil. Soc.} {\bf#1}, #2 (#3). }
\def\pl#1#2#3   {{ Phys. Lett.} {\bf#1}, #2 (#3). }
\def\plc#1#2#3   {{ Phys. Lett.} {\bf#1}, #2 (#3); }
\def\prep#1#2#3 {{ Phys. Rep.} {\bf#1}, #2 (#3). }
\def\prev#1#2#3 {{ Phys. Rev.} {\bf#1}, #2 (#3). }
\def\prl#1#2#3  {{ Phys. Rev. Lett.} {\bf#1}, #2 (#3). }
\def\prs#1#2#3  {{ Proc. Roy. Soc.} {\bf#1}, #2 (#3). }
\def\ptp#1#2#3  {{ Prog. Th. Phys.} {\bf#1}, #2 (#3). }
\def\rmp#1#2#3  {{ Rev. Mod. Phys.} {\bf#1}, #2 (#3). }
\def\rpp#1#2#3  {{ Rep. Prog. Phys.} {\bf#1}, #2 (#3). }
\def\zp#1#2#3   {{ Zeit. Phys.} {\bf#1}, #2 (#3). }
\def\epj#1#2#3   {{ Eur. Phys. Jour.} {\bf#1}, #2 (#3). }
\def\nim#1#2#3   {{ Nucl. Instr. Meth.} {\bf#1}, #2 (#3). }

\newcommand{\rmt}{\rm\textstyle}
\newcommand{\rms}{\rm\scriptstyle}
\newcommand{\stw}{\mbox{$\sin^2\theta_W$}}
\newcommand{\stwos}{\mbox{$\sin^2\theta_W^{\rms(on-shell)}$}}
\newcommand{\nub}{\overline{\nu}}

\newcommand{\alphabar}[0]{\mbox{$\overline{\alpha}$}}
\newcommand{\kappabar}[0]{\mbox{$\overline{\kappa}$}}
\newcommand{\qbar}[0]{\overline{q}}

\newcommand{\sbar}[0]{\overline{s}}

\newcommand{\Qbar}[0]{\overline{Q}}
\newcommand{\Ubar}[0]{\overline{U}}

\newcommand{\Sbar}[0]{\overline{S}}
\newcommand{\Dbar}[0]{\overline{D}}
\newcommand{\uubar}[0]{\mbox{$\stackrel{(-)}{u}$}}
\newcommand{\ddbar}[0]{\mbox{$\stackrel{(-)}{d}$}}

\newcommand{\uavbar}[0]{\langle\overline{u}(x)\rangle}
\newcommand{\cavbar}[0]{\langle\overline{c}(x)\rangle}
\newcommand{\savbar}[0]{\langle\overline{s}(x)\rangle}
\newcommand{\davbar}[0]{\langle\overline{d}(x)\rangle}

\newcommand{\Uavbar}[0]{\langle\overline{U}\rangle}
\newcommand{\Cavbar}[0]{\langle\overline{C}\rangle}
\newcommand{\Savbar}[0]{\langle\overline{S}\rangle}
\newcommand{\Davbar}[0]{\langle\overline{D}\rangle}
\newcommand{\qav}[0]{\langle{q(x)}\rangle}
\newcommand{\uav}[0]{\langle{u(x)}\rangle}
\newcommand{\cav}[0]{\langle{c(x)}\rangle}
\newcommand{\sav}[0]{\langle{s(x)}\rangle}
\newcommand{\dav}[0]{\langle{d(x)}\rangle}
\newcommand{\Qav}[0]{\langle{Q}\rangle}
\newcommand{\Uav}[0]{\langle{U}\rangle}
\newcommand{\Cav}[0]{\langle{C}\rangle}
\newcommand{\Sav}[0]{\langle{S}\rangle}
\newcommand{\Dav}[0]{\langle{D}\rangle}
\newcommand{\nubar}[0]{\overline{\nu}}

\newcommand{\nunub}{\stackrel{{\footnotesize (-)}}{\nu}}
\newcommand{\txnunub}[1]{\nu_{#1},\nub_{#1}}

\newcommand{\Rnu}{\mbox{$R^{\nu}$}}
\newcommand{\Rnub}{\mbox{$R^{\nub}$}}

\begin{document}

\preprint{UR-1647}
\preprint{FNAL-Pub-02/050-E}

\title{On the Effect of Asymmetric Strange Seas and Isospin-Violating Parton
  Distribution Functions on $\boldmath\stw$ Measured in the NuTeV Experiment}

\author{G.~P.~Zeller$^{5}$, K.~S.~McFarland$^{8,3}$,   
 T.~Adams$^{4}$, A.~Alton$^{4}$, S.~Avvakumov$^{8}$, 
 L.~de~Barbaro$^{5}$, P.~de~Barbaro$^{8}$, R.~H.~Bernstein$^{3}$, 
 A.~Bodek$^{8}$, T.~Bolton$^{4}$, J.~Brau$^{6}$, D.~Buchholz$^{5}$, 
 H.~Budd$^{8}$, L.~Bugel$^{3}$, J.~Conrad$^{2}$, R.~B.~Drucker$^{6}$, 
 B.~T.~Fleming$^{2}$, R.~Frey$^{6}$, J.A.~Formaggio$^{2}$, J.~Goldman$^{4}$, 
 M.~Goncharov$^{4}$, D.~A.~Harris$^{8}$, R.~A.~Johnson$^{1}$, J.~H.~Kim$^{2}$,
 S.~Koutsoliotas$^{2}$, M.~J.~Lamm$^{3}$, W.~Marsh$^{3}$, D.~Mason$^{6}$, 
 J.~McDonald$^{7}$, C.~McNulty$^{2}$, 
    D.~Naples$^{7}$, 
 P.~Nienaber$^{3}$, A.~Romosan$^{2}$, W.~K.~Sakumoto$^{8}$, H.~Schellman$^{5}$,
 M.~H.~Shaevitz$^{2}$, P.~Spentzouris$^{2}$, E.~G.~Stern$^{2}$, 
 N.~Suwonjandee$^{1}$, M.~Tzanov$^{7}$, M.~Vakili$^{1}$, A.~Vaitaitis$^{2}$, 
 U.~K.~Yang$^{8}$, J.~Yu$^{3}$, and E.~D.~Zimmerman$^{2}$}
\affiliation{
$^1$University of Cincinnati, Cincinnati, OH 45221 \\
$^2$Columbia University, New York, NY 10027 \\
$^3$Fermi National Accelerator Laboratory, Batavia, IL 60510 \\
$^4$Kansas State University, Manhattan, KS 66506 \\
$^5$Northwestern University, Evanston, IL 60208 \\
$^6$University of Oregon, Eugene, OR 97403 \\
$^7$University of Pittsburgh, Pittsburgh, PA 15260 \\
$^8$University of Rochester, Rochester, NY 14627 \\ 
}
\date{\today}
\begin{abstract}
  
  The NuTeV collaboration recently reported a value of $\stw$ measured in
  neutrino-nucleon scattering that is 3 standard deviations above the
  standard model prediction.  This result is derived assuming
  that (1) the strange sea is quark-antiquark symmetric, $s(x)=\sbar(x)$, and
  (2) up and down quark distributions are symmetric under the
  simultaneous interchange of $u\leftrightarrow d$ and $p\leftrightarrow n$.
  We report the impact of violations of these symmetries on $\stw$ and
  discuss the theoretical and experimental constraints on 
  such asymmetries.

\end{abstract} 

\maketitle

%

\section{Introduction and Formalism}

Based on
measurements of neutral current and charged current neutrino-nucleon 
scattering in both neutrino and anti-neutrino beams, 
the NuTeV collaboration recently reported a measurement of $\stwos$. 
The result \cite{nc-prl},
\begin{eqnarray}
    \sin^2\theta_W^{({\rms on-shell)}}&=&0.2277\pm0.0013({\rmt stat.})\pm0.0009(
    {\rmt syst.})
        \nonumber\\
        &-&0.00022\times(\frac{M_{top}^2-(175 \: \mathrm{GeV})^2}{(50 \: \mathrm
{GeV})^2})
        \nonumber\\
        &+&0.00032\times \ln(\frac{M_{Higgs}}{150 \: \mathrm{GeV}}),
\end{eqnarray}
\noindent
is approximately 3 standard deviations above the expected value of 
$0.2227\pm0.0004$ \cite{LEPEWWG,Martin}.

Ratios of
neutral current to charged current cross sections on isoscalar targets
of $u$ and $d$ quarks are experimental observables
that can be related to fundamental electroweak parameters. 
Before NuTeV, high statistics 
neutrino experiments measured $\stw$ using the Llewellyn Smith 
cross section ratios \cite{llewellyn}:
\begin{equation}
R^{\nu(\nub)} \equiv \frac{\sigma(\nunub N\rightarrow\nunub X)}
                 {\sigma(\nunub N\rightarrow\ell^{-(+)}X)}  
= g_L^2+r^{(-1)}g_R^2,
\label{eqn:ls}
\end{equation}
where
\begin{equation}
r \equiv \frac{\sigma({\overline \nu}N\rightarrow\ell^+X)}
                {\sigma(\nu N\rightarrow\ell^-X)} \sim \frac{1}{2},  
\label{eqn:rdef} 
\end{equation}
and 
\begin{eqnarray}
g_L^2 & = & (\epsilon^u_L)^2+(\epsilon^d_L)^2 \nonumber\\
          &  = & \frac{1}{2}-\stw+\frac{5}{9}\sin^4\theta_W, \nonumber\\
g_R^2 & = & (\epsilon^u_R)^2+(\epsilon^d_R)^2 \nonumber\\
          &  = & \frac{5}{9}\sin^4\theta_W 
                 .
\end{eqnarray}
\noindent
For the experimental values of $r$ and $\stw$, it follows that
$\Rnu$ is much more sensitive to $\stw$ than is $\Rnub$. 

Inspired by the Paschos-Wolfenstein relationship \cite{Paschos-Wolfenstein}:
\begin{eqnarray}
R^{-} &\equiv& \frac{\sigma(\nu_{\mu}N\rightarrow\nu_{\mu}X)-
                   \sigma(\nub_{\mu}N\rightarrow\nub_{\mu}X)}
                  {\sigma(\nu_{\mu}N\rightarrow\mu^-X)-  
                   \sigma(\nub_{\mu}N\rightarrow\mu^+X)} \nonumber\\  
&=& \frac{\Rnu-r\Rnub}{1-r} = g_L^2-g_R^2,
\label{eqn:rminus}
\end{eqnarray}
\noindent
NuTeV uses high statistics separated neutrino and anti-neutrino beams to
measure $\stw$ and 
thereby reduces its sensitivity to uncertainties in cross sections
resulting from scattering off $q$-$\qbar$ symmetric quark seas. Using
the separate neutrino and antineutrino data sets, NuTeV also extracts 
effective neutral current quark couplings, $(g_L^{\rms eff})^2$ and 
$(g_R^{\rms eff})^2$ \cite{nc-prl}.

Let $\qav$ denote the momentum distribution of a particular flavor of quark
averaged over the nucleons in the NuTeV target, and let $\Qav \equiv \int
\qav dx$, the total momentum carried by quark flavor $q$.  Let
nucleon-specific quark momentum distributions be denoted by $q_p(x)$ and
$q_n(x)$, with corresponding integrals $Q_p$ and $Q_n$, respectively.  Both
the Llewellyn Smith and Paschos-Wolfenstein relationships assume $\Uav=\Dav$
and $\Uavbar=\Davbar$.  The Llewellyn Smith interpretation of $\Rnu$
assumes additionally that $\Sav=\Savbar=\Cav=\Cavbar$ (clearly $\Sav=\Cav$ is
experimentally not a good assumption), while the Paschos-Wolfenstein $R^-$
formula assumes only $\Sav=\Savbar$ and $\Cav=\Cavbar$.

The NuTeV $\stw$ analysis accounts for the violations of the assumption that
$\uav=\dav$ and $\uavbar=\davbar$ which result from the excess of neutrons
over protons in the target.  From a material inventory of the NuTeV target
calorimeter, we measure a 
$5.74\pm0.02\%$ 
fractional excess of neutrons over
protons \cite{King-thesis}.  However, the NuTeV result assumes exact isospin
symmetry in neutron and proton quark distributions,
$\uubar_p(x)=\ddbar_n(x)$, $\ddbar_p(x)=\uubar_n(x)$. The NuTeV analysis
assumes furthermore that $\sav=\savbar$ and $\cav=\cavbar$.  It has been
pointed out that such assumptions, if incorrect, produce sizable shifts in
the NuTeV $\stw$\cite{Sather,Thomas,Cao,Gambino}.

Although the NuTeV experiment does not exactly measure $R^-$, in part because
it is not possible experimentally to measure neutral current reactions down
to zero recoil energy, it is nevertheless illustrative to calculate the effect 
of these violations on $R^-$.  Denote the neutron excess of the NuTeV target 
as
$\delta N \equiv (A-2Z)/A$
and the total valence momentum carried by the
proton as $V_p=U_p-\Ubar_p+D_p-\Dbar_p$.  Let the following
\begin{eqnarray}
\delta D_v & \equiv &   D_p-\Dbar_p-U_n+\Ubar_n \nonumber\\
\delta U_v & \equiv &   U_p-\Ubar_p-D_n+\Dbar_n \nonumber\\
\delta \Dbar & \equiv &  \Dbar_p-\Ubar_n \nonumber\\
\delta \Ubar & \equiv &  \Ubar_p-\Dbar_n \nonumber\\
\delta S & \equiv &   \Sav-\Savbar
\end{eqnarray}
denote deviations from the above symmetry assumptions.   To first order 
in $\delta N$, $\delta Q_v$, $\delta\Qbar$ and $\delta S$, we obtain
\begin{eqnarray}
R^- & \approx & \Delta_u^2+\Delta_d^2 \nonumber\\
&& - \: \delta N \left( \frac{U_v-D_v}{V_p}\right) 
                    (3\Delta_u^2+\Delta_d^2) \nonumber\\
&& + \: \frac{\delta U_v-\delta D_v}{2V_p}  
                    (3\Delta_u^2+\Delta_d^2)  \nonumber\\
&& + \: \frac{\delta S}{V_p} (2\Delta_d^2-3(\Delta_d^2+\Delta_u^2)\epsilon_c),  
\label{eqn:deltaR-}
\end{eqnarray}
where $\Delta_{u,d}^2 = (\epsilon^{u,d}_L)^2-(\epsilon^{u,d}_R)^2$ and where
$\epsilon_c$ denotes the ratio of the scattering cross section from the
strange sea including kinematic suppression of heavy charm production
to that without
kinematic suppression.  In this calculation, we assume the massless 
quark-parton model which implies no longitudinal cross section, no target 
mass effects, and we also assume $\Cav=\Cavbar=0$.

As already noted, to extract $\stw$, NuTeV does not measure directly $R^-$,
but rather measures ratios of experimental candidates within kinematic
criteria and compares this to a full Monte Carlo simulation which accounts
for neutral current and charged current cross-talk, non-quark-parton model
contributions to the cross section, radiative corrections, electron neutrino
backgrounds, and detector resolution \cite{nc-prl}.  Therefore, the NuTeV
$\stw$ measurement does not depend on these symmetry violating terms in
the way that Equations~\ref{eqn:rminus} and \ref{eqn:deltaR-} would
suggest.

\newcommand{\eE}{\mbox{$\cal{E}$}}
To examine the exact effect of various symmetry violations on the 
NuTeV analysis, we first define a functional $F[\eE,\delta; x]$ such that the 
shift in an experimental quantity, $\eE$, due to a symmetry violating
quark fractional momentum distribution, $\delta(x)$, is given by:

\begin{equation}
\Delta\eE = \int_0^1 F[\eE,\delta;x]\,\delta(x)\:dx.
\end{equation}

\begin{figure}
\epsfxsize=0.47\textwidth\epsfbox{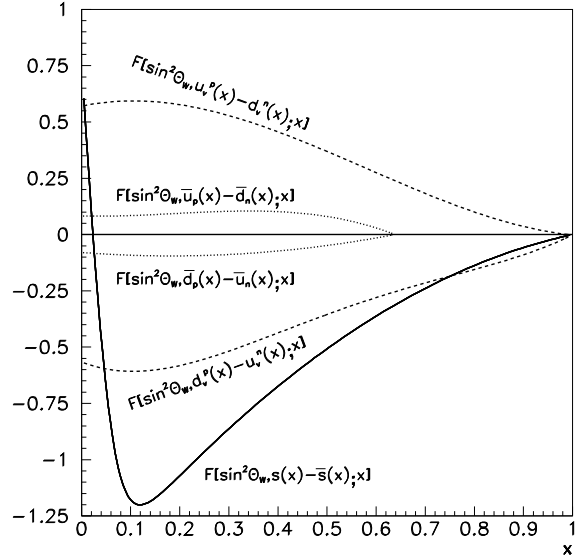}
\caption{The functionals describing the shift in the NuTeV $\stw$ caused by
  not correcting the NuTeV analysis for isospin violating $u$ and $d$
  valence and sea distributions or for $\sav\neq\savbar$.  The shift in
  $\stw$ is determined by convolving the asymmetric momentum distribution with
  the plotted functional.}
\label{fig:stw}
\end{figure}

\begin{figure}
\epsfxsize=0.47\textwidth\epsfbox{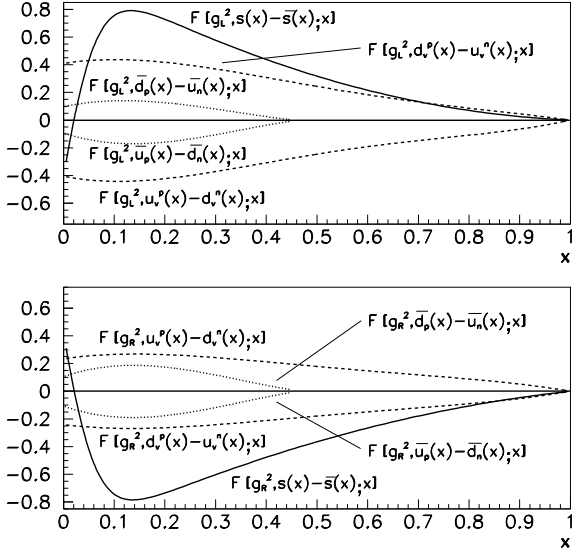}
\caption{The functionals describing the shifts in the NuTeV 
  $(g_L^{\rms eff})^2$ and $(g_R^{\rms eff})^2$ caused by not correcting 
  the NuTeV analysis for isospin violating $u$ and $d$ valence and sea 
  distributions or for $\sav\neq\savbar$. The shifts in
  $(g_L^{\rms eff})^2$ and $(g_R^{\rms eff})^2$ are 
  determined by convolving the asymmetric momentum distribution with
  the plotted functional.}
\label{fig:glgr}
\end{figure}

\noindent
All of the details of the NuTeV Monte Carlo simulation and 
measurement can be parameterized in terms of $F[\eE,\delta;x]$, and therefore,
this formalism provides a way to determine the shift in the NuTeV
measurement for arbitrary symmetry violation in PDFs. Figures~\ref{fig:stw}
and \ref{fig:glgr} show $F[\eE,\delta;x]$ for an isospin symmetry
violating $u$ and $d$ valence and sea and for $\sav\neq\savbar$.
Figure~\ref{fig:stw} shows the functionals for the NuTeV measurement of $\stw$,
while Figure~\ref{fig:glgr} shows the corresponding functionals for
$(g_L^{\rms eff})^2$ and $(g_R^{\rms eff})^2$.

\section{Asymmetric Strange Sea}

If the strange sea is generated by purely perturbative QCD processes, then 
neglecting electromagnetic effects, one expects $\sav=\savbar$.  However, it 
has been noted that non-perturbative QCD effects can
generate a significant momentum asymmetry between the strange and anti-strange 
seas \cite{Signal,Burkardt,Brodsky,Melnit}. Lending weight to this possibility, a 
joint fit to CDHS neutrino charged-current inclusive cross
sections \cite{cdhs} 
(but not including CCFR \cite{unki} and NuTeV data 
or neutrino dimuon cross sections) and charged lepton 
structure function data reports some improvement in their fits if they allow 
for an asymmetry in the strange sea at high $x$ \cite{Barone}.  
The CCFR and CDHS charged current neutrino
cross-sections differ significantly at high $x$ where this joint fit
finds a large strange sea asymmetry, $s\gg\sbar$.

By measuring the 
processes $\txnunub N\to \mu^+\mu^- X$
the CCFR and NuTeV experiments constrain the difference between the
momentum distributions of the strange and anti-strange seas. For studying the
effect on the NuTeV $\stw$, it is important to 
study such effects within the same PDF formalism and corresponding 
cross sections as were used in the measurement itself \cite{nc-prl}.
In this enhanced leading order cross section model, the CCFR/NuTeV $\nu, 
\nubar$ dimuon data were fit \cite{max} to the following form 
for the strange and anti-strange seas \footnote{At $Q^2=16$~GeV$^2$, the
  average $Q^2$ of the NuTeV data used in the \stw\ analysis, the NuTeV 
  $(\uavbar+\davbar)/2$ can be parameterized as
  $e^{-0.75-150x}+e^{-1.33-7.7x-8.1x^2}$ over the region $0<x<0.6$.  
  NuTeV determines its leading
  order PDFs from fits to CCFR cross section data including external
  constraints \cite{nc-prl,unki}.}:
\begin{eqnarray}
\sav&=&\kappa\frac{\uavbar+\davbar}{2}(1-x)^{\alpha} \nonumber\\
\savbar&=&\kappabar\frac{\uavbar+\davbar}{2}(1-x)^{\alphabar}, 
\label{eqn:ssbar}
\end{eqnarray}
\noindent
obtaining central values of
\begin{equation}
 \left( \begin{array}{c} \kappa \\ \kappabar \\ \alpha \\ \alphabar
       \end{array}\right) 
=
 \left( \begin{array}{c} .352 \\ .405 \\ -0.77 \\ -2.04
       \end{array}\right) 
\end{equation}
and a covariance matrix\footnote{This covariance matrix is from the fit of
  Ref.~\onlinecite{max}, although the matrix is not given in the original
  paper.}  incorporating both statistical and systematic uncertainties on
these parameters:
\begin{equation}
 \left( 
       \begin{array}{cccc}
          0.0034 &  0.0027 & -0.028 & -0.007 \\
          0.0027 &  0.0031 & -0.024 & -0.008 \\
         -0.028  & -0.024  &  0.78  &  0.18 \\
         -0.007  & -0.008  &  0.18  &  0.29 
       \end{array}
\right) .
\end{equation}
Within this particular model, the measurement implies a {\em negative} 
asymmetry,
\begin{equation}
 \Sav-\Savbar = -0.0027 \pm 0.0013,
\end{equation}
\noindent
and a resulting increase in the NuTeV value of $\stw$,
\begin{equation}
 \Delta\stw = +0.0020 \pm 0.0009.
\end{equation}
\noindent
The initial NuTeV measurement, which assumes $\sav=\savbar$, becomes 
$\stw$ = $0.2297 \pm 0.0019$. Hence, if we use the experimental measurement
of the strange sea asymmetry, the discrepancy
with the standard model is increased to $3.7\sigma$ significance.

A recent calculation \cite{Gambino} claims that a {\em
positive} strange sea asymmetry of $\Sav-\Savbar = +0.0020$ could
explain half of the NuTeV discrepancy ($\Delta\stw = -0.0026$).  It
should be noted, however, that this is an overestimate, 
as Figure~\ref{fig:stw}
makes clear, due to the fact that charged current charm suppression threshold 
effects have been neglected in their analysis, and because NuTeV does not 
exactly measure $R^-$ \footnote{The inclusion of an asymmetric strange sea 
induces a larger and opposite sign shift in $\Rnub$ compared to the shift 
in $\Rnu$. Because the NuTeV result is less sensitive to $\Rnub$ than is
$R^-$, the effect is reduced at all $x$.  The large suppression of
charged-current scattering from the low $x$ strange sea explains the change of
sign in the shift in $\stw$ at very low $x$.}.

Reference \cite{Barone} reports favoring a significant
positive strange sea asymmetry ($S-\Sbar\sim+0.0020$) at high $x$. A fit to
the form assumed in Equation~\ref{eqn:ssbar} does not necessarily 
exclude such an
asymmetry as it is dominated by data at low $x$. The asymmetry of
Reference \cite{Barone} would imply at least a 5$\%$ increase in the total
$\nu$ dimuon cross section in the region $x>0.5$. However, NuTeV has
looked for such an excess at high $x$ and excludes additional
dimuon sources larger than 0.2$\%$ (0.6$\%$) in the $\nu$ ($\nub$) data at
90$\%$ confidence \cite{max}.

\section{Isospin Violating PDFs}

Several recent classes of non-perturbative models predict 
isospin violation in the nucleon \cite{Sather,Thomas,Cao}. We
evaluate the shift in the NuTeV value of $\stw$ under the assumption that the
asymmetry occurs in nature and is not corrected for in the NuTeV
analysis.  The earliest estimation in the literature, a bag model calculation
\cite{Sather}, predicts large valence asymmetries of opposite sign in
$u_p-d_n$ and $d_p-u_n$ at all $x$, which would produce a shift in
the NuTeV $\stw$ of $-0.0020$.  However, this estimate neglects a number of
effects, and a complete calculation by Thomas {\em et al.\,} \cite{Thomas}
concludes that asymmetries at very high $x$ are larger, but the asymmetries 
at moderate $x$ are smaller and of opposite sign at low $x$, thereby 
reducing the shift in $\stw$ to a negligible $-0.0001$.  Finally, the effect 
is also evaluated in the Meson Cloud model \cite{Cao}, and there the 
asymmetries are much smaller at all $x$, resulting in a modest shift in the 
NuTeV $\stw$ of $+0.0002$.

The calculation of Thomas {\em et al.\,} \cite{Thomas} is particularly useful 
in evaluating uncertainties because it
decomposes isospin violating effects into different parts that are driven by
experimental or theoretical inputs.  The largest contributions to a shift in
$\stw$ in this calculation come from the single quark ($m_d-m_u\sim4$~MeV)
and nucleon ($m_n-m_p\approx1.29$~MeV) mass differences.  The former has a
significant theoretical uncertainty, and we assign a fractional error of
$25\%$ to this source of isospin violation based on the uncertainty in
$m_d-m_u$\cite{Leut,Bickerstaff}; such an uncertainty translates to a
$0.0001$ uncertainty in the NuTeV $\stw$.  Another contribution 
in this calculation with large
theoretical uncertainties is the effect of diquark
($m_{dd}-m_{uu}$) mass differences. This causes isospin breaking
predominantly at high $x$ where both the PDFs are small {\em and} the effect
on the NuTeV measurement is negligible. The uncertainty is therefore 
significantly smaller than that from the single quark mass shift.

In general, nuclear effects can also cause isospin-breaking, thereby
producing $\Uav\neq \Dav$ in the NuTeV target, which is primarily iron.  
While less theoretically certain, one estimate of the effect 
exists \cite{Davidson} and would predict a modest increase in the NuTeV $\stw$.

Although a particular nucleon or nuclear charge symmetry violation model could
account for the NuTeV discrepancy with the standard model, such models,
in their attempt to explain the NuTeV $\stw$, must 
be evaluated in the context of a global fit to all experimental data derived 
from any such asymmetry assumptions because they may disagree with 
existing data \cite{bodek}.

\section{Conclusions}
The fact that NuTeV does not measure directly $R^-$ or exact ratios of 
neutral to charged current cross sections makes it difficult to 
predict the effect of parton level symmetry violations. Hence, 
we present a framework for evaluating the effects of both isospin violating 
$u$ and $d$ parton densities and asymmetric strange seas on the NuTeV 
measurements of $\stw$, $(g_L^{\rms eff})^2$, and $(g_R^{\rms eff})^2$.
While it is possible, in principle, to induce sizable shifts in the NuTeV
$\stw$ with variations in the former, the joint CCFR/NuTeV neutrino and
anti-neutrino dimuon data limit possible charge asymmetry in the
strange sea. In fact, relaxing the restriction that $\sav=\savbar$ in the
LO fit to CCFR/NuTeV dimuon data increases the NuTeV discrepancy 
with the standard model.

\begin{acknowledgements}
This work was supported by the U.S. Department of 
Energy, the National Science Foundation, and the Alfred P. Sloan Foundation.
\end{acknowledgements}
%

\end{document}